\documentclass[mathleft,fleqn,%
]{an}
\usepackage{graphicx}
\usepackage[varg]{txfonts}
\usepackage{natbib}
\bibpunct{(}{)}{;}{a}{}{,}
\begin{document}

\Pagespan{1}{}
\Yearpublication{2014}%
\Yearsubmission{2014}%
\Month{0}%
\Volume{999}%
\Issue{0}%
\DOI{asna.201400000}%

\title{Near-UV transit photometry of HAT-P-32\,b with the LBT\thanks{Based on observations made with the Large Binocular Telescope (LBT). The LBT is an international collaboration among institutions in the United States, Italy and Germany. LBT Corporation partners are: LBT Beteiligungsgesellschaft, Germany, representing The Leibniz Institute for Astrophysics Potsdam, the Max-Planck Society, and Heidelberg University; The University of Arizona on behalf of the Arizona Board of Regents; Istituto Nazionale di Astrofisica, Italy; The Ohio State University, and The Research Corporation, on behalf of The University of Notre Dame, University of Minnesota and University of Virginia.}: Silicate aerosols in the planetary atmosphere}

\author{Matthias Mallonn\inst{1}\fnmsep\thanks{Correspondence:
        {mmallonn@aip.de}}
\and Hannah\,R. Wakeford\inst{2}
}
\titlerunning{Near-UV transit photometry of HAT-P-32\,b}
\authorrunning{M. Mallonn \& H.R. Wakeford}
\institute{
Leibniz Institute for Astrophysics Potsdam (AIP), An der Sternwarte 16, 14482 Potsdam, Germany
\and 
Planetary Systems Laboratory, NASA Goddard Space Flight Center, Greenbelt, MD 20771, USA
}


\received{XXXX}
\accepted{XXXX}
\publonline{XXXX}


\abstract{Broad-band exoplanet transit photometry can characterize the planetary atmosphere when observed at multiple selected filters. This observing technique can reveal gradients in the spectra of extrasolar planets, for example the slope of decreasing opacity from short to long optical wavelengths caused by aerosol scattering. In this work we observed a transit of the hot Jupiter HAT-P-32\,b in the shortest wavelength possible from the ground using the Large Binocular Telescope (LBT). The data comprise the best-quality ground-based U-band taken so far of an exoplanet transit. Compared to broad-band observations of intermediate and long optical wavelength published previously, a clear scattering slope in the planetary transmission spectrum is revealed. Most likely, the scattering particles are magnesium silicate aerosols larger than 0.1 $\mu$m. We define a spectral index to compare this scattering feature of HAT-P-32\,b to published results of other exoplanets. It turns out to be very typical in amplitude within the comparison sample. Furthermore, we searched for correlation in this sample of the spectral index with planetary equilibrium temperature, surface acceleration and stellar activity indicator, but could not reveal any. }

\maketitle

\section{Introduction}
The majority of known host stars of transiting extrasolar planets are solar-type stars, i.\,e. main sequence stars of spectral type late F, G, and K. These stars emit only little flux at ultraviolet (UV) wavelengths, hence UV observations of transit events often suffer from photon noise and do not reach the same very high quality as we are used to nowadays for optical transit photometry \citep{Nascimbeni2013,Mallonn2015,Turner2016}. This is an unfortunate situation since the UV wavelengths are key for at least two open questions in exoplanet science. One concerns the interaction of close-in gas giants with the stellar corona and the potential formation of a bow shock in front of the planet, which might be detectable in absorption in the UV \citep{Vidotto2011}. The second open question is on the presence of a scattering signature in the transmission spectra of extrasolar planets, which would be most pronounced at the bluest wavelengths \citep[e.\,g.,][]{Lecavelier2008,Fortney2010,Wakeford2015,Mallonn2016}.

WASP-12\,b was observed by the Hubble Space Telescope (HST) to show an asymmetric transit light curve at near-UV wavelengths \citep[254--258~nm,][]{Fossati2010}. The transit started significantly earlier at these short wavelengths than its counterparts at optical wavelengths. This early ingress was confirmed by follow-up observations of WASP-12\,b at 278.9--282.9~nm \citep{Haswell2012}. Two main models for explaining the origin of an early ingress were given, both including the formation of a bow-shock, one of magnetic origin \citep{Vidotto2011,Vidotto2010}, the other of non-magnetic origin \citep{Lai2010}. Recently, \cite{Turner2016} observed a large sample of hot Jupiter transit events in the Johnson U band to investigate whether the phenomenon of an early transit ingress is common among close-in gas giants. The precision of their measurements were sufficient to rule out similar transit asymmetries as in the HST measurements of WASP-12\,b. Thus, the observation of bow-shock phenomena might be constrained to wavelengths shortward of 300~nm.

Cloud- and haze-free atmospheres of extrasolar planets are expected to show an increased opacity towards blue optical wavelengths and the near-UV because of light scattering by H$_2$ molecules \citep{Fortney2010}. Also small-sized haze particles (aerosols) cause scattering with a dominant slope in the near-UV and optical transmission spectrum \citep[e.\,g.,][]{Pont2013,Sing2013,Nikolov2015}. This increased opacity by scattering is measurable as an increased effective planetary radius and therefore increased transit depth during transit compared to longer wavelength. In a large and homogeneous study, \cite{Sing2016} analyzed ten close-in gas giants and found in all cases an enlarged planetary radius at bluest wavelength. The amplitude can be compared among different planets in terms of their atmospheric pressure scale height. All system showed an increase in effective planetary radius, however with diverse amplitudes \citep[note also the conflicting result for HAT-P-12\,b by][]{Mallonn2015}. Thus, is the blue scattering signature a spectral signature common to all transmission spectra of close-in gas giants? This question can be answered by high-quality transit light curves taken at very short wavelengths like the Johnson U band at 360~nm. There are numerous examples of ground-based U band transit observations in the literature \citep{Nascimbeni2013,Copperwheat2013,Bento2014,Turner2016,Turner2013,Mallonn2016,Mallonn2015,Kirk2016}, however the derived value of planetary size rarely reaches a precision of two scale heights or below. 

Whether a variation in planetary size of only a few scale heights is measurable as a variation in transit depth depends on the planet-star radius ratio and the size of the scale height. Additionally, also the brightness of the host star matters to lower the photon noise in the transit photometry. One very favorable exoplanet is HAT-P-32\,b with a fairly bright F-type host star of V\,=\,11.4~mag, an inflated planetary radius of $\sim$1.7~$R_J$ and a scale height of $\sim$1000~km \citep{Hartman2011}. Several publications exist on the optical transmission spectrum of HAT-P-32\,b \citep{Gibson2013,MallonnStrass,Nortmann2016,Mallonn2016}, and they all agree on the absence of pressure-broadened absorption features predicted by cloud-free atmosphere models. However, the current data bluer than about 450~nm could not definitively clarify about a scattering signature because of a decrease in precision towards these short wavelengths. In what follows, we present the analysis of a near-UV broad-band observation of a transit of HAT-P-32\,b observed with the LBT. In Section 2 we describe the observation and data reduction, while in Section 3 the analysis and results are presented. In Section 4 we discuss the broad-band spectrum of HAT-P-32\,b in regard to literature spectra of other close-in exoplanets. Section 5 contains the conclusion.

\section{Observation and data reduction}

\begin{figure*}
\includegraphics[height=\hsize,angle=270]{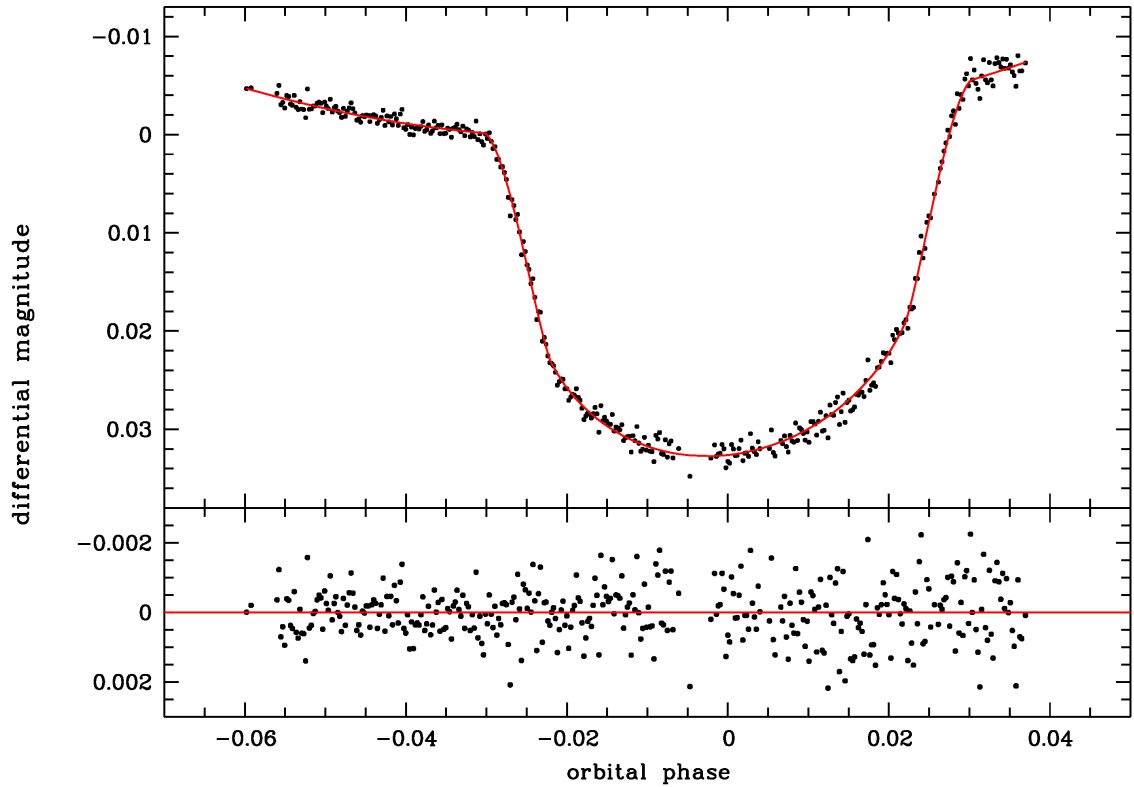}
\caption{U band transit light curve of HAT-P-32\,b. The red solid line shows the best fit model. Below, the residuals are plotted.}
\label{plot_lc}
\end{figure*}

We observed one transit with the Large Binocular Camera \citep[LBC,][]{Giallongo2008} at the Large Binocular Telescope (LBT) on November 29, 2016. The LBC consists of two prime-focus, wide-field imagers mounted on the left and right arm of LBT, and is optimized for blue and red optical wavelengths, respectively. The observation was performed with the LBT in binocular mode using the blue optimized LBC on one side, and the Multi-Object Double Spectrograph (MODS) on the other side. The analysis of the MODS data is work in progress and is not included in this paper. The LBC already proved its capabilities for exoplanet transit and secondary eclipse observations \citep{Nascimbeni2013,Mallonn2015,vonEssen2015}. We employed the U$_{spec}$ filter, which has a response curve very similar to Sloan u' but increased efficiency. The exposure time was 15~s, and we reduced the overheads by a read-out window to 29~s. The transit lasted from November 30, 07:03 to 10:09~UT, and the observation was executed from 05:25 to 10:25~UT in which we gathered 375 science images. The object was setting over the course of the observation, thus the airmass increased towards the end to 2.25. The telescope was heavily defocused to avoid detector saturation.

The data reduction was done as described in our previous work on HAT-P-32\,b \citep{Mallonn2016}. Bias and flat-field correction was done in the standard way, with the bias value extracted from the overscan regions. We performed aperture photometry with the publicly available software SExtractor using the option of a fixed circular aperture MAG\_APER and the automatically adjusted aperture MAG\_AUTO. The set of comparison stars (flux sum) was chosen to minimize the root mean square (rms) of the light curve residuals after subtraction of a second order polynomial over time plus transit model using literature transit parameter. Using the same criterion of a minimized rms, we also determined and applied the best aperture width to be MAG\_AUTO with a Kron parameter of 2.75 (see the SExtractor manual\footnote{http://www.astromatic.net/software/sextractor} for details). The best circular aperture with a diameter of 72 pixel gave a slightly larger point-to-point scatter. 

\section{Light curve analysis and results}

The transit light curve analysis was performed homogeneously to our previous studies on the same target \citep{MallonnStrass,Mallonn2016} to allow for a direct comparison. We model the transit light curve with the publicly available software JKTEBOP \citep{Southworth2004,Southworth2008} in version 34. The transit fit parameters consist of the sum of the fractional planetary and stellar radius, $r_{\star} + r_p$, and their ratio $k=r_p/r_{\star}$, the orbital inclination $i$, the transit midtime $T_0$, the host-star limb-darkening coefficients (LDC) $u$ and $v$ of the quadratic limb darkening law, and the coefficients $c_{0,1,2}$ of a polynomial over time. The index ``$\star$'' refers to the host star and ``p'' refers to the planet. The dimensionless fractional radius is the absolute radius in units of the orbital semi-major axis $a$, $r_{\star} = R_{\star}/a$, and $r_p = R_p/a$. The planetary eccentricity is fixed to zero as determined by \cite{Zhao2014} and the orbital period $P_{\mathrm{orb}}$ is fixed to 2.15000825 days according to \cite{Seeliger2014}.

The stellar limb darkening was approximated by the quadratic law \citep[see][for a comparison of different limb darkening laws for HAT-P-32]{Mallonn2016}. Theoretical values of the LDC were taken from \cite{Claret2013}, we used the stellar parameters of the host star derived by \cite{Hartman2011} for the circular orbit solution. Homogeneously to \cite{Mallonn2016}, we fitted the linear LDC $u$ and kept $v$ fixed to its theoretical value.

\begin{figure}
\includegraphics[height=\hsize,angle=270]{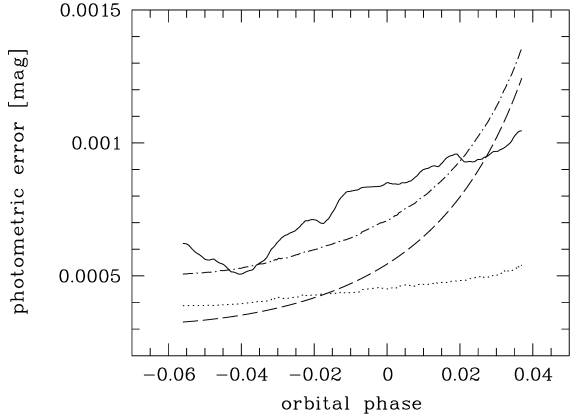}
\caption{Photometric uncertainty of the individual data points of the time series. The theoretical value of photon noise of target, comparison stars, skybackground and read-out noise is given as dotted line, the theoretical scintillation noise as dashed line, and their quadratically combined value as dashed-dotted line. The solid line is the measured uncertainty as running standard deviation.}
\label{plot_err}
\end{figure}

After an initial run of transit modeling, we noticed an increase in point-to-point scatter with time for the light curve residuals. This increase is not reflected in the SExtractor calculation of the photometric uncertainties that takes into account the photon noise of the target and the measured noise in the local background. We quadratically combined the uncertainties of the target and the comparison stars. In the first part of our time-series, the point-to-point scatter is approximately 1.3 above the theoretical SExtractor uncertainty, while towards the end it is larger by more than a factor of two. \cite{Dravins1997} noticed that scintillation noise caused by the Earth atmosphere strengthens towards shorter wavelengths, thus we calculated the scintillation noise with the approximating Equation 10 of \cite{Dravins1998} (Fig. \ref{plot_err}). And indeed, during the second half of our observations the scintillation noise is much larger than the photon noise. In an attempt to use the most realistic photometric uncertainties in our light curve modeling, we derived the standard deviation per data point in the light curve residuals with a sliding window of 60 data points width (a ``running standard deviation'' in analogy to a running mean) and applied this as uncertainty. As shown in Figure \ref{plot_err}, this photometric uncertainty varies from $\sim$\,0.5 to $\sim$\,1.0~mmag (time sampling of 44 seconds) with a mean of 0.78~mmag, which is to our knowledge the best ground-based near-UV transit light published so far.

Additionally, we calculated the so-called $\beta$ factor, a concept introduced by \cite{Gillon06} and \cite{Winn08} to include the contribution of correlated noise in the light curve analysis. It describes the evolution of the standard deviation $\sigma $ of the light curve residuals when they become binned in comparison to Poisson noise. In the presence of correlated noise, $\sigma $ of the binned residuals is larger by the factor $\beta$ than with pure uncorrelated (white) noise. The value of $\beta\,=\,1.15$ derived for our light curve is the average of the values for a binning from 10 to 30 minutes in two minute steps. We enlarged the photometric uncertainty by this factor.

A smooth trend in the out-of-transit baseline is obvious in Fig. \ref{plot_lc}. We tried to model this trend with a linear combination of linear functions of airmass, detector position, width of the point-spread-function, and sky background \citep{vonEssen2016}, but in our case none of the terms was verified by the Bayesian Information Criterion \citep[BIC,][]{Schwarz78}. We found a simple second-order polynomial over time to approximate the trend very well and to minimize the BIC.

As a consistency check to previous studies on HAT-P-32\,b we fitted our near-UV transit light curve with all transit and detrending parameters free. Our derived values of $i\,=\,88.7\pm0.7$ and $a/R_{\star}\,=\,6.09\pm0.27$ agree to within $1\,\sigma$, while the $k\,=\,0.1548\pm0.0011$ is larger than literature values by 2 to $3\,\sigma$. The transit midtime $T_0\,=\,2\,457\,722.859898\pm0.00012$ BJD$_{\mathrm{TDB}}$ is in $1\,\sigma$ agreement to the ephemeris of \cite{Seeliger2014}. The estimation of the transit parameter uncertainties was done with  ``task 8'' in \mbox{JKTEBOP} \citep{Southworth2005}, which is a Monte Carlo simulation, and with ``task 9'' \citep{Southworth2008}, which is a residual-permutation algorithm that takes correlated noise into account. We run the Monte Carlo simulation with 5000 steps. As final parameter uncertainties we adopted the larger value of both methods.

\begin{figure}
\includegraphics[height=\hsize,angle=270]{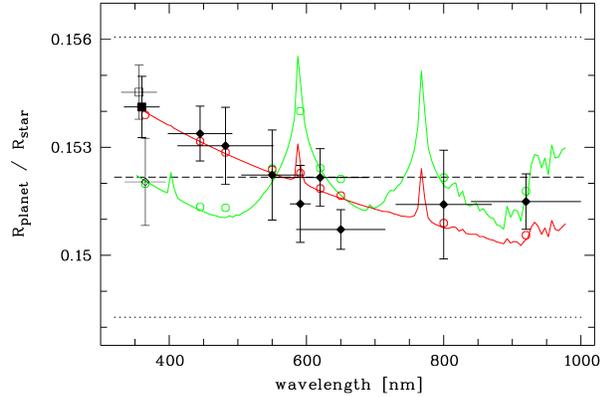}
\caption{Broad-band transmission spectrum of HAT-P-32\,b. Measurements of \cite{Mallonn2016} use open and filled diamond symbols. The U band data point of this work is given as open square, the result of the joint analysis of the two U band light curve of this work and \cite{Mallonn2016} is presented as filled square. The green model represents a cloud-free atmosphere with solar-composition and TiO/VO removed from \cite{Fortney2010}, and the red model includes additionally a Rayleigh-scattering source. The dashed horizontal line gives the weighted average of the filled symbol data points. The dotted horizontal lines are the $\pm$3~scale height values. }
\label{plot_transm}
\end{figure}

We want to investigate the wavelength dependence of the effective planetary radius. We assume $a/R_{\star}$ and $i$ to be wavelength independent and fix them to the values used in \cite{MallonnStrass} and \cite{Mallonn2016} for comparability. We assume their uncertainties to be a common source of noise to all bandpasses, negligible in the search for relative variations of $k$ over wavelength. Therefore, the derived error on $k$ is a relative uncertainty. The free parameters of the light curve fit are $r_p$, $u$, and $c_{0,1,2}$. The LDC $v$ is fixed to its theoretical value as done in \cite{MallonnStrass} and \cite{Mallonn2016}.
The result is shown in Fig. \ref{plot_transm} in comparison to \cite{Mallonn2016}, who derived a U band data point from a single ULTRACAM transit light curve. We performed a joint fit of both U band light curves with JKTEBOP, allowing for an individual detrending of both light curves, but a common value for $r_p$ and $u$. The results are summarized in Tab. \ref{tab_k}.

\begin{table*}
\caption{Derived transit parameters of this work in comparison to literature values for HAT-P-32\,b. The limb darkening coefficient $u$ was fitted in the analysis, while the coefficient $v$ was fixed to its theoretical value.}
\label{tab_k}
\begin{center}
\begin{tabular}{lccccc}\hline
Date & Telescope &  $k$ & $u$ & $v$ & reference \\
\hline
2012 Oct 12 & WHT & 0.1520 $\pm$ 0.0012 & 0.543 $\pm$ 0.023 & 0.112 & \cite{Mallonn2016} \\
2016 Nov 29 & LBT & 0.15453 $\pm$ 0.00075 & 0.565 $\pm$ 0.015 & 0.112 & this work \\
\hline
\multicolumn{2}{c}{joint analysis} & 0.15412 $\pm$ 0.00085 & 0.571 $\pm$ 0.019 & 0.112 & \\

\hline
\end{tabular}
\end{center}
\end{table*}

\section{Discussion}

\subsection{No early ingress -- a symmetric transit light curve}

Our data yield no indication of an asymmetry in the transit light curve or an early ingress, potentially caused by a bow shock in the stellar corona as for the transit observations of WASP-12\,b shortward of 300~nm \citep{Fossati2010,Haswell2012}. The transit depth of HAT-P-32\,b in the U band is only slightly larger than at optical wavelengths without indications of absorption in a planetary exosphere extending its Roche lobe. With this result we are in agreement to the observational result of \cite{Turner2016}, who did not find an asymmetry or early ingress for 15 different hot Jupiters using ground-based U band observations. The study of the origin of these transit asymmetries appears to be possible only at UV wavelengths from space.

\subsection{Aerosol scattering in the planetary atmosphere}

The analysis of the U band light curve of HAT-P-32\,b resulted in an increased planet-star radius ratio $k$ compared to longer wavelength measurements of \cite{Mallonn2016}. Under the assumption of a stellar radius independent of wavelength, it translates into an increased near-UV effective radius of the planet. The parameter $k$ could in principle also be modified by third light from another stellar body falling in the photometric aperture. And indeed, \cite{Adams2013} found a near-by M dwarf companion only 2.9\,$''$ to HAT-P-32. However, the third light contribution of this M dwarf towards the total system light at near-UV wavelengths is smaller than 0.1\,\%, and therefore the modification in $k$ is smaller by orders of magnitude than its uncertainty \citep{MallonnStrass}. In principle, also brightness inhomogeneities on the stellar surface (star spots) could cause variations in $k$ \citep{Oshagh2014,Herrero2016}. However, \cite{MallonnStrass} found the host to be photometrically stable, therefore star spots are very unlikely to cause a significant increase in $k$ towards blue wavelengths. 

The most plausible explanation for a larger effective radius of the planet at near-UV wavelengths is scattering in the planetary atmosphere. The scattering particles could be hydrogen molecules in gas phase \citep{Fortney2010}. However, this scenario is unlikely since it would require a cloud/haze-free atmosphere, which is disfavored by all previous spectroscopic studies of this planet \citep{Gibson2013,MallonnStrass,Nortmann2016}. Certainly, the scattering takes place at aerosols, which are thought to be a widespread phenomenon in the atmospheres of hot Jupiter exoplanets \citep{Lodders2010,Marley2013}. In a recent work, \cite{Barstow2017} found all of the 10 planets of the HST spectral survey \citep{Sing2016} to contain aerosols at the probed latitudes, however at varying pressure levels.

\begin{figure}
\includegraphics[width=\hsize]{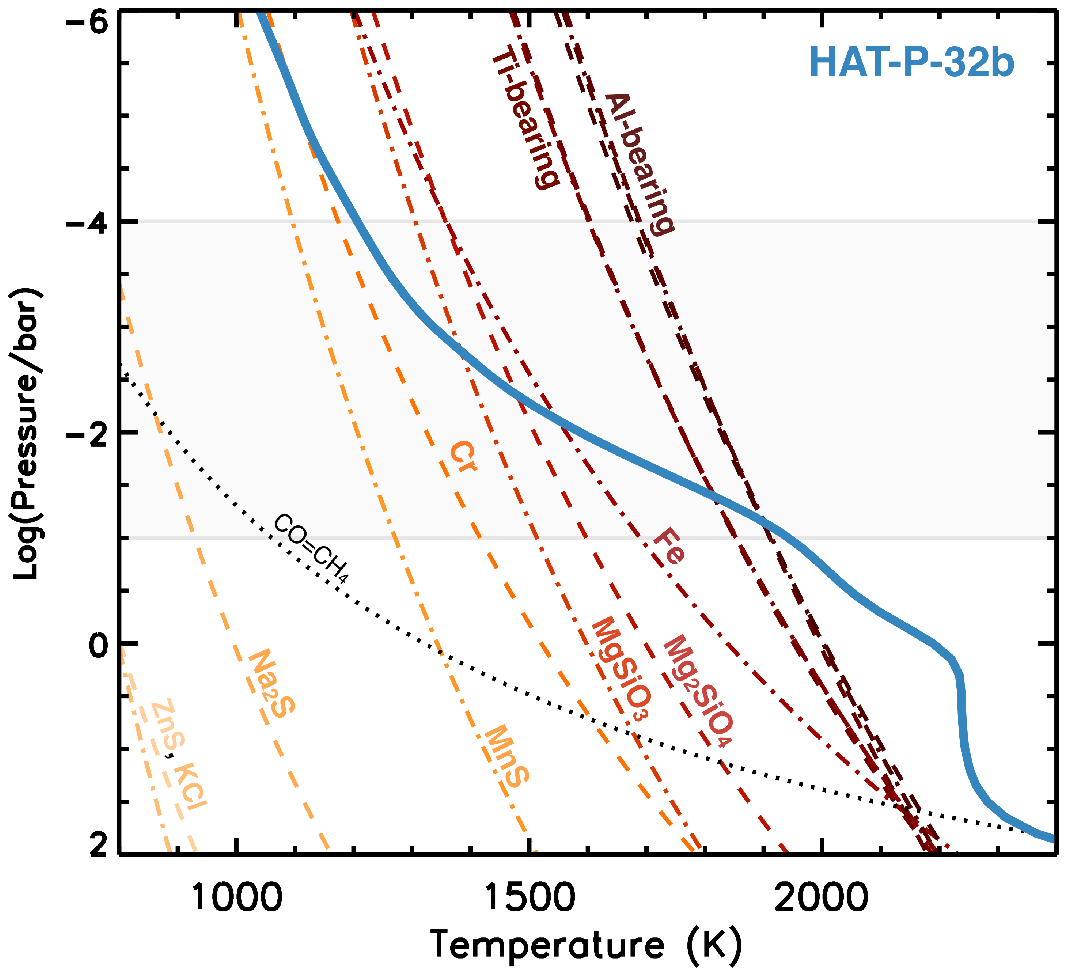}
\caption{Globally-averaged T-P profile for HAT-P-32b (blue) with T$_{eq}$\,=\,1753\,K, a gravity of 7\,ms$^{-2}$, and solar metallicity without TiO, calculated following the formulation in \citet{Fortney2008,Fortney2013}. We show the condensation curves for relevant cloud species detailed in \citet{Visscher2010, Morley2012, Wakeford2017}. Where the T-P profile crosses the condensation curves indicates the base of the condensate cloud. The region within the two horizontal lines shows the approximate pressures probed transmission spectral observations, these U band observations are likely in the top range of these pressures compared to near-infrared features. We also indicate the gas phase transition from CO to CH$_4$ showing that HAT-P-32b clouds are not expected to be photochemical in nature (e.g. \citealt{Moses2011}).}
\label{plot_cond}
\end{figure}

We show the temperature-pressure (T-P) profile of HAT-P-32b in Fig. \ref{plot_cond} with the pressures probed by transmission spectral measurements constrained by vertical lines to indicate the region of the atmosphere measured through observations. It is expected that the U band measurements presented in this paper are probing the top regions of the atmosphere at higher altitudes than would be observed in the near-IR, however, further constraints cannot be placed on the T-P profile. We use a series of condensation curves (\citealt{Visscher2010, Morley2012, Wakeford2017}) to approximate the different cloud forming species expected in the upper atmosphere of HAT-P-32b. These condensation curves indicate that the most likely cloud forming species at the pressures probed in transmission would be composed of magnesium silicates, which have been hypothesized as sources of enhanced optical scattering in a number of hot Jupiters (e.g. \citealt{Wakeford2015,Nikolov2015}) and are expected to be abundant in giant exoplanet atmospheres (e.g. \citealt{Lodders2010, Visscher2010}).

In an attempt to put our new HAT-P-32\,b optical transmission spectrum in the context of the existent observations of other exoplanets, we defined the spectral index $\Delta k_{u-z}$. It measures the difference in $k$ between U band wavelengths (here defined as $300<\lambda<420$~nm) and z' band wavelengths (here defined as $880<\lambda<1000$~nm) and expresses it in terms of the atmospheric pressure scale height $H$ for reasonable comparability among different planets. The scale height $H$ is defined by
\begin{equation}
H\,=\,\frac{k_B\,T_{\mathrm{eq}}}{\mu_m\,g}
\end{equation}
with $T_{\mathrm{eq}}$ as the planetary equilibrium temperature, $k_B$ the Boltzmann constant, $\mu_m$ the mean molecular mass equaling 2.2, and g as the local gravitational acceleration.
For the calculation of $H$ we used published values of $T_{\mathrm{eq}}$ and $g$ (see Tab. \ref{tab_para}). The index $\Delta k_{u-z}$ can only be calculated for planets for which the U and z' band planet-star radius ratios have been derived consistently, i.e. with a common value for $i$ and $a/R_{\star}$, and the use of the same limb darkening law. Together with the values for HAT-P-32\,b of this work, we could therefore include nine planets investigated by \cite{Sing2016}, the results of GJ3470\,b from \cite{Nascimbeni2013}, and an independent measurement of HAT-P-12\,b by \cite{Mallonn2015}, a planet also included in the sample of \cite{Sing2016}. If the wavelength interval used here for U and z' band are spectroscopically resolved by multiple data points, we use a weighted average.

Under the assumption of the equilibrium temperature being a correct description of the average temperature at the terminator, the value $\Delta k_{u-z}$ can be used as an indicator for the particle size of the scattering aerosols. Particles of the size 0.1~$\mu$m and smaller cause a steep Rayleigh scattering slope, while larger particles cause a shallower slope. The transition to a flat spectrum happens for particle sizes larger than $\sim$\,0.5 to 1~$\mu$m with a slight dependence on atmospheric temperature \citep{Wakeford2015,Wakeford2017}. 

The optical spectral slope $\Delta k_{u-z}$ of HAT-P-32\,b of $2.18\,\pm\,0.95$~$H$ is slightly smaller than the expectation of $\sim$\,4~$H$ for a pure Rayleigh-scattering slope. In comparison to the other investigated systems in Tab. \ref{tab_para}, the value for HAT-P-32\,b appears to be about average. A majority of targets show a $\Delta k_{u-z}$ of between one and three $H$. On the contrary, the results of $\Delta k_{u-z}\,>\,5~H$ of HD189733b and GJ3470b are unusually large. This favors a usually larger atmospheric particle size for most of the hot Jupiters than for these two planets, i.\,e. larger than 0.1~$\mu$m, but smaller than $\sim$\,0.5~$\mu$m. Another valid explanation for an optical slope shallower than expected for Rayleigh-scattering could be a difference in the scale height of the aerosols to that of the surrounding gas. The value $\Delta k_{u-z}\,\sim\,5$~H of HD189733\,b is indicative for Rayleigh scattering by aerosols with a similar scale height as the gas, while the lower value of $\Delta k_{u-z}\,\sim\,2$~H for, e.\,g., WASP-12\,b could potentially be explained by Rayleigh scattering of aerosols with a lower scale height than the gas \citep{Sing2013}. One way to measure the aerosol scale height might be the detection of vibrational mode absorption features of the aerosols at mid-infrared wavelengths with JWST \citep{Wakeford2015,Wakeford2017} and a comparison of their amplitudes with theoretical models.

\begin{figure}
\includegraphics[width=\hsize]{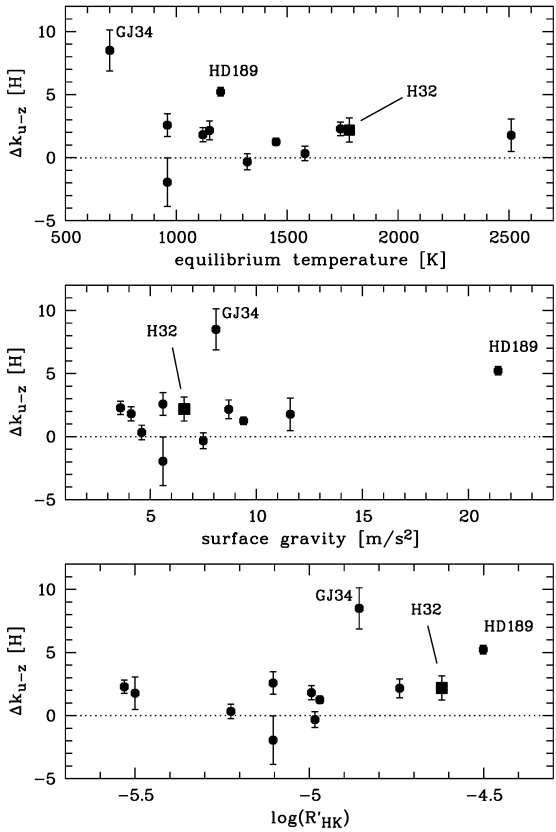}
\caption{Spectral index $\Delta k_{u-z}$ in relation to planetary and stellar parameter. The upper plot shows the spectral index over the planetary equilibrium temperature $T_{\mathrm{eq}}$, the middle plot the index in regard to the planetary surface gravity $g$, and the lower panel the index in regard to the stellar activity indicator log(R'$_{\mathrm{HK}}$). HAT-P-32\,b is indicated as squared data point versus the dotted data points of the comparison sample.}
\label{plot_para}
\end{figure}

The temperature of hot Jupiters ranges from below 1000~K to 3000~K or above. The elemental composition of a cloud is expected to vary with temperature because of different condensation temperature for the different potential condensation species. We searched for a correlation of the spectral index $\Delta k_{u-z}$, potentially being related to the particle size, with $T_{\mathrm{eq}}$, but none was found (upper panel of Fig. \ref{plot_para}). The coolest planet in the sample, GJ3470b, has the strongest optical spectral slope. However, all other planets follow a roughly constant value of $\Delta k_{u-z}$ over $T_{\mathrm{eq}}$. We also searched for a relation of $\Delta k_{u-z}$ and the planetary surface gravity $g$, because for larger $g$ values, larger particles might have been settled below observable altitudes. Thus, we might expect a positive correlation of $\Delta k_{u-z}$ and $g$. We find HD189733b with its largest $g$ value among our sample to show a large $\Delta k_{u-z}$ value. However, all other planets do not show a significant correlation (middle panel of Fig. \ref{plot_para}). Another test, which we performed, looked for a correlation among $\Delta k_{u-z}$ and the stellar activity indicator log(R'$_{\mathrm{HK}}$). The aerosols in the planetary atmospheres might be formed by condensation or photochemistry. The latter is expected to be important for cooler hot Jupiter atmospheres, while the former might dominate the hotter atmospheres \citep{Wakeford2015}. Photochemistry is linked to the incident UV flux that the planet receives. To quantify this, we use the indicator log(R'$_{\mathrm{HK}}$), since stellar activity correlates with enhanced UV flux \citep{Knutson2010}. Hydrocarbon species formed by photochemistry are expected to be small in particle size \citep{Wakeford2015}, therefore we would intuitively expect a positive correlation between $\Delta k_{u-z}$ and log(R'$_{\mathrm{HK}}$). However, the planets of our sample form a wide range in log(R'$_{\mathrm{HK}}$), and except HD189733b having an extreme value of log(R'$_{\mathrm{HK}}$) and a large value of $\Delta k_{u-z}$, there is no indication for a dependence.

\begin{table*}
\caption{The difference in planet-star radius ratio $\Delta k_{u-z}$ in units of H. The physical parameters have been obtained by \cite{Sing2016}, \cite{Astudillo2016}, \cite{Hartman2011} and the online database www.exoplanet.eu.}
\label{tab_para}
\begin{center}
\begin{tabular}{lccccccc}\hline
Object & $T_{\mathrm{eq}}$ [K] &  $g$ [m/s$^2$] & log(R'$_{\mathrm{HK}}$) & $H$ [km] & $\Delta k_{u-z}$ [$H$] & Reference \\
\hline
GJ3470b    &  700   &   8.1  &  -4.85  &  326   &  8.50  $\pm$  1.63  &  \cite{Nascimbeni2013}  \\
HAT-P-12b  &  960   &   5.6  &  -5.10  &  648   &  2.58  $\pm$  0.90  &  \cite{Sing2016} \\
HAT-P-12b  &  960   &   5.6  &  -5.10  &  648   &  -1.94  $\pm$  1.92  &  \cite{Mallonn2015} \\
WASP-39b   &  1120  &   4.1  &  -4.99  &  1032  &  1.81  $\pm$  0.56  &  \cite{Sing2016} \\
WASP-6b    &  1150  &   8.7  &  -4.74  &  499   &  2.15  $\pm$  0.74  &  \cite{Sing2016} \\
HD189733b  &  1200  &   21.4 &  -4.50  &  212   &  5.22  $\pm$  0.33  &  \cite{Sing2016} \\
HAT-P-1b   &  1320  &   7.5  &  -4.98  &  665   &  -0.32  $\pm$  0.62  &  \cite{Sing2016} \\
HD209458b  &  1450  &   9.4  &  -4.97  &  583   &  1.25  $\pm$  0.28  &  \cite{Sing2016} \\
WASP-31b   &  1580  &   4.6  &  -5.22  &  1298  &  0.33  $\pm$  0.57  &  \cite{Sing2016} \\
WASP-17b   &  1740  &   3.6  &  -5.53  &  1827  &  2.28  $\pm$  0.52  &  \cite{Sing2016} \\
HAT-P-32b  &  1780  &   6.6  &  -4.62  &  1019  &  2.18  $\pm$  0.95  &  this work \\
WASP-12b   &  2510  &   11.6 &  -5.50  &  818   &  1.77  $\pm$  1.28  &  \cite{Sing2016} \\
\hline
\end{tabular}
\end{center}
\end{table*}

\section{Conclusion}

We observed one transit of the inflated hot Jupiter HAT-P-32\,b with the LBT in the U band. With an average point-to-point scatter (rms) of 0.78~mmag and a time sampling of 44~seconds, these data form the best-quality ground-based near-UV transit light curve so far. We do not find indication of an asymmetry or an early ingress in the transit light curve, indicating that studies on bow shock phenomena in the stellar corona or on the exosphere of hot Jupiters are bound to wavelengths shortward of 300~nm. Such observations cannot be obtained from the ground. We reveal an increase in planet-star radius ratio in the U band compared to optical wavelengths previously published. A plausible explanation is scattering in the planetary atmosphere caused by aerosols. By a comparison of the planetary T-P profile with condensation curves of potential cloud forming species we find the aerosols to be most likely formed by magnesium silicates. The scattering signature in the planetary broad-band transmission spectrum is smaller than expected for Rayleigh-scattering, which is indicative for an aerosol particle size larger than 0.1 up to 0.5~$\mu$m. 

We defined the spectral index $\Delta k_{u-z}$, which measures the difference in the planet-star radius ratio between the U band and the z' band in units of the planetary atmospheric pressure scale height $H$. This index allowed a comparison of the scattering feature derived for HAT-P-32\,b in this work with published results of other exoplanets. We find that the scattering feature of HAT-P-32\,b is rather typical for hot Jupiter systems. More pronounced scattering features like in the atmosphere of HD189733\,b are rather an exception. We searched for correlation of $\Delta k_{u-z}$ with the planetary equilibrium temperature, the planetary surface gravity and the stellar magnetic activity indicator, but did not find any. Our sample of 11 targets with measurements of $\Delta k_{u-z}$ might still be too small. We also urge the community to observe targets more widespread in the $T_{\mathrm{eq}}$--$g$ parameter space to reveal dependencies of the optical spectral scattering slope. 

\acknowledgements
We thank Jonathan J. Fortney for providing the planetary atmosphere models and the planetary T-P profile. We also like to thank Jesper Storm and the LBT Science Operation Team for performing the observations at the LBT and the helpful discussions regarding their preparation. This research has made use of the SIMBAD data base and VizieR catalog access tool, operated at CDS, Strasbourg, France, and of the NASA Astrophysics Data System (ADS).

\bibliographystyle{an}
\bibliography{H32_U_bib}

\end{document}